\documentclass[aip,amsmath,amssymb,reprint]{revtex4-1}

\usepackage{color}
\usepackage{graphicx}
\usepackage{dcolumn}% Align table columns on decimal point
\usepackage{bm}% bold math
\usepackage[utf8]{inputenc}
\usepackage[T1]{fontenc}
\usepackage{mathptmx}
\usepackage{etoolbox}

\makeatletter
\def\@email#1#2{%
 \endgroup
 \patchcmd{\titleblock@produce}
  {\frontmatter@RRAPformat}
  {\frontmatter@RRAPformat{\produce@RRAP{*#1\href{mailto:#2}{#2}}}\frontmatter@RRAPformat}
  {}{}
}%
\makeatother

\def\<{\langle}
\def\>{\rangle}
\def\[{\left[}
\def\]{\right]}
\def\({\left(}
\def\){\right)}

\def\re{ {\rm Re} }
\def\im{ {\rm Im} }

\def\Eh{~$E_{\rm h}$}

\def\a0{~$a_0$}

\def\MM#1#2#3#4{\left[ {\begin{array}{cc} {#1} & {#2} \\ {#3} & {#4} \end{array} } \right] }
\def\MMF#1#2#3#4#5#6#7{\left[ {\begin{array}{cccc} 
	{#1} & {#5} & {#6} & {#7} \\
	{#5} & {#2} & {} & {} \\
	{#6} & {} & {#3} & {} \\
	{#7} & {} & {} & {#4} 
\end{array} } \right] }

\def\au{a.u.}
\def\mylabel#1{\label{#1}}

\def\Ltwo{${\cal L}^2$~}

\begin{document}

\preprint{AIP/123-QED}

\title{
	Complex scaling spectrum using multiple avoided crossings at stabilization graph
}

\author{Petra Ruth Kapr\'alov\'a-\v Z\v d\'ansk\'a}
\email{kapralova@fzu.cz}
\affiliation{
{$^1$ Department of Radiation and Chemical Physics,
Institute of Physics, Academy of Sciences of the Czech Republic,
Na Slovance 2, 182 21 Prague 8, Czech Republic}
}

\date{\today}% It is always \today, today,

\begin{abstract}
This study concerns finite basis set $\{\chi_k\}$ calculations of
resonances based on real scaling, $\chi_k(x)\to \chi_k(xe^{-\eta})$.
I demonstrate that resonance width
is generally influenced by {\it several} neighboring quasi-discrete continuum states.
Based on this finding I propose a new method to calculate the complex
resonance energy together with several states of complex rotated continuum.
The theory is introduced for a one-dimensional model, then
it is applied for helium doubly excited resonance $2s^2$.
The new method requires the real spectrum (``stabilization graph'') 
for a sufficiently large interval of the parameter $\eta$
on which the potential curve of the sought resonance gradually meets
several different quasi-continuum states.
Diabatic Hamiltonian which comprehends the resonance and the several quasi-continuum states 
participating at the avoided crossings is constructed.
As $\eta$ is taken to complex plane, $\eta\to i\theta$, the
corresponding part of the complex scaled spectrum is obtained. 
\end{abstract}

\maketitle

\section{Introduction}

\noindent
Resonance state represents quantum particle which is temporarily
trapped within a confined space therefore
resonance wavefunction resembles bound state wavefunction
with an outgoing wave~\cite{Kapralova-Zdanska:2011}.
The norm of the captured quantum particle
 drops down in time according to the first order kinetics,
which inherently brings about a diverging character of the
outgoing wave and also explains complex energy eigenvalue
of resonance state with imaginary part given by
half decay rate~\cite{Kapralova15}.
Wherefore resonances do not belong to \Ltwo space.

Different computational approaches to calculate resonances  (auto-ionizing states) 
have been developed.
Perhaps the most famous method is represented by {\it complex scaling}~\cite{Reinhardt1982,MoiseyevRev3}
and  {\it exterior complex scaling}~\cite{Simon1979,McCurdy2004,McCurdy1978,Moiseyev1979}
of Hamiltonian (or interchangeably basis set),
which effectively transforms resonances onto \Ltwo space.
These methods enable to construct non-Hermitian Hamiltonian 
which provides complex energies of resonances and rotated continuum.
%(I pointed out some specific numerical properties of the complex scaling method
%in~\cite{Kapralova15} and proposed a physical interpretation of the complex solutions in~\cite{Kapralova21}.)

Another approach to resonances is represented by {\it stabilization methods}.
These methods are often used for ab initio calculations of molecular
resonances as their main advantage is that they rely on Hermitian calculations.
Resonances are derived from real energies obtained for states 
above ionization threshold, namely from {\it stabilization graph}, 
which is represented by above-threshold potential energy curves obtained as the
problem is parametrized in various ways, e.g. real scaling of the basis set~\cite{Simons1981,Thompson1982,McCurdy1983b},
or adding a binding potential~\cite{Chao1990,Horacek2010,Curik2016}.
Complex resonance energies are obtained via extrapolation of the
free parameter to complex plane. 

Yet another method, popular for its relatively simple implementation
to quantum chemistry packages, is represented by 
{\it complex absorbing potential}~\cite{Jolicard1985,Jolicard1986,Riss1983,Riss1995}. 
Basic idea here is enforcing outgoing boundary conditions via additional
artificial imaginary potential term which is localized in the asymptotic region.

Apart of these three types of methods, there are also approaches based on
projection operators~\cite{Hazi1978},
scattering theory~\cite{Morrison1977,Rescigno1974I,Rescigno1974II,Langhoff1976},
or Siegert state expansion~\cite{Tolstikhin2008}
available to calculate resonances.

In this paper I propose a new method with the unique feature to
calculate partial complex scaling spectrum, 
{\it resonances and rotated continuum}, using
stabilization graph. 
The ability to calculate discretized rotated continuum 
is important for a subsequent implementation of strong field interactions~\cite{ISI:A1994NR28400030,ISI:A1994PV95600060,ISI:A1994NL68500030}.
Future intended use of the proposed method is particularly
for ab initio simulations of laser-atom interactions phenomena such as 
high-order harmonics generation using standard quantum chemistry packages.
The paper is organized as follows.

First, in Section~II, I discuss the relationship between stabilization method,
complex coordinate scaling, and complex absorbing potential, where the argument is
backed up by illustrative calculations for a one-dimensional model potential.
It is shown that avoided crossings found at stabilization
graph continue to complex plane of the free parameter, where they end up in exceptional
points (EPs)~\cite{McCurdy1983b}. Such EPs
 designate transition between adiabatic
and diabatic regimes where resonance decouples from quasi-continuum states.

In Section~III, I suggest a diabatic Hamiltonian
for a {\it set} of avoided crossings in the stabilization graph, 
where the diabatic states include the resonance state defined
by constant real energy,
and several close states of discretized continuum,
which are analytically dependent on the free real parameter (here
I use real scaling parameter of the basis set $\chi_k(x)\to\chi_k(e^{-\eta})$).
Their analytical dependence is derived from the known behavior of
the free particle confined in a large box.
The new diabatic Hamiltonian is complex scaled by taking
the parameter $\eta$ to complex plane, $\eta\to i\theta$, 
which provides partial complex scaling spectrum, namely the complex
resonance energy and few states of the complex rotated continuum.
It is shown that the obtained complex energies
correspond one-to-one with a direct application of
the complex scaling method when using the same system and
basis set.

In Section~IV, the new method is implemented for a typical  ab initio
calculation of atomic resonances, which is represented by the
$2s^2$ resonance of the helium atom where a large-scale basis set
ExTG5G\cite{Kapralova-Zdanska:2013a} is used.
The method is fine-tuned for this purpose by adding some improvements
such as a precise diabatization method, improved 
analytical form for discretized continuum states due to the specific basis set,
extrapolation and correction methods for the complex resonance energy.   
I conclude that the method is robust when using quasi-complete
Gaussian basis set, yielding unique results.

In Section~V, the present theoretical findings are summarized, 
pros and cons of the proposed method are discussed,
 and prospective applications are suggested.

\section{Theory}

\subsection{Methods based on a scaling of the box size}

\noindent
Let me start with a discussion of logical connections between the
methods of complex scaling, stabilization graph, and complex absorbing potential.

First, the idea of the stabilization method is achieving a variable box size, for
which different strategies are used.
The most straighforward way is represented by scaling 
a finite basis set $\{\chi(x)\}$ directly (for a reason that will be explained later we
choose the exponential form of the scaling parameter), $\{\chi(x)\}\to \{\chi(xe^{-\eta})\}$, $\eta\in \Re$. 
The box size is defined indirectly by the finite phase space, which
is associated with the \Ltwo basis set used, and effectively varied by the scaling. 
Adding a binding potential to the physical Hamiltonian has the same effect.
For example, a binding
Coulomb term was added to an electronic Hamiltonian to get stabilization
graphs for molecules~\cite{Horacek2010}; namely, the ionized states in the true system
were replaced by the Coulomb states of the added Coulomb cone (functioning as the ``box'')
and as the artificial ``charge'' was varied,
the width of the Coulomb cone (the ``box'') has been effectively changed.

The real scaling of the basis set turns to complex scaling of the
Hamiltonian if only the parameter $\eta$ is taken imaginary. This is
the basic idea of the stabilization method; an analytical
continuation of energy to complex plane of $\eta$ is usually done 
using Pad\'e approximants.
If the real potential is added, the principle of analytical continuation
is applicable supposed a removal of the non-physical potential
via extrapolation.

In the complex absorbing potential method, an imaginary term ($-iV(x), V(x)>0$)
is added near edges of the phase-space area covered by the basis set ($x>x_0$).
This idea, based on quantum dynamics, is that the imaginary term suppresses outgoing wavefunctions;
resonances, including outgoing but not incomming wave, thus become part of
the \Ltwo space.
One can find here a connection to the coordinate scaling, too.
Let me give an example of adding a real quadratic barrier, $V(x>x_0) = \eta^2 (x-x_0)^2$.
A variation of $\eta$ represents a real scaling of the outer part of the basis set, 
namely for $x>x_0$, and would allow for a construction of a stabilization graph. 
The complex absorbing potential is obtained as $\eta^2$ is analytically continued 
to complex plane such that $\eta^2\to-|\eta|^2\,\sqrt{i}$.

\subsection{Decoupling of a resonance and quasi-continuum at an exceptional point}

\noindent
Energy spectrum using any type of scalable box size allows one to
distinguish between different types of energy levels which are among the states of quasi-continuum.
But let me start with the bound states by pointing out that their energies are not affected by the scaling parameter,
as one scales only the outer box but not the inner (e.g. nuclear) potential.
The first type of the quasi-continuum states are the quasi-free states, where the particle is out of the physical potential.
The quasi-free spectrum obviously depends on the box size.
The second type of the quasi-continuum states are the quasi-bound states, where the particle is temporarily trapped in the 
physical potential, however, it might be released via the tunneling phenomenon or others.
Now, the presence of the outer box creates an artificial situation where
the quasi-bound particle is forced to stay bound, unless there exists a quasi-free state with a similar energy 
to which it can couple.
Such a situation is
intentionally created by varying the scaling parameter (whether $\eta$, or the charge in the case of an additional Coulomb
potential, etc.). 
The obtained picture is of course the stabilization graph, where
the spectrum is plotted in the form of potential energy curves as functions of the scaling parameter.
The graph includes regions of stability, where the quasi-bound state appears as a constant energy curve,
disrupted by avoided crossings as a quasi-free state approaches its energy.

Let me introduce a simple example to illustrate the ideas just mentioned -- a one-dimensional model potential,
\begin{align}
V = -v_0\, e^{-\frac{x^2}{\sigma_0^2}} + v_1\[\,e^{-\frac{(x-x_0)^2}{\sigma_1^2}} + e^{-\frac{(x+x_0)^2}{\sigma_1^2}}\],
\mylabel{eqV}
\end{align}
where $v_0 = 7.1$~\au, $v_1 = 4.5$~\au, $\sigma_0 = 4$~\au, $\sigma_1 = 2$~\au,
 which
supports bound states and shape-type resonances,  Fig.~1.
By scaling the basis set with the real scaling parameter $\exp(-\eta)$,
which is here represented by the box states
\begin{align}
\chi_n(x;\eta) = L_{\eta}^{-1/2}\,\sin\( \frac{x + L_\eta}{2\,L_\eta}\,{n\pi}\) ,\quad L_\eta = L_0\,e^\eta, 
\mylabel{eqbas}
\end{align} 
one obtains the stabilization graph shown in Fig.~2.
\begin{figure}
	\includegraphics[width=3in]{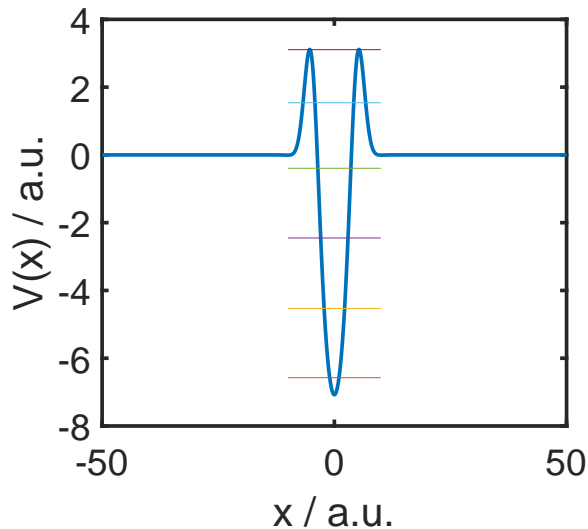}
	\caption{Model one-dimensional potential, Eq.~\ref{eqV}, used for present demonstrations, supports bound states and resonances shown by the energy levels.}
	\mylabel{F1}
\end{figure}
\begin{figure}
	\includegraphics[width=3in]{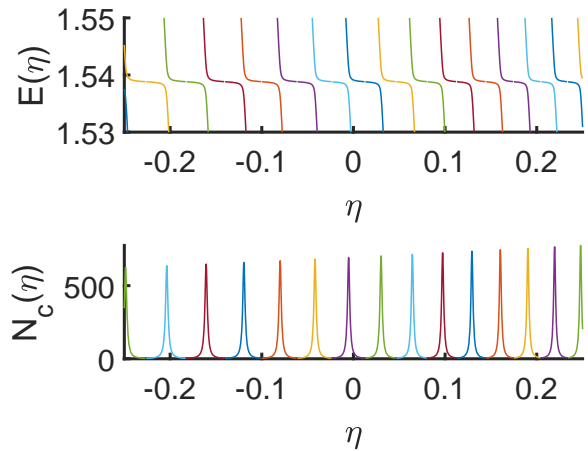}
	\caption{
	(a) Stabilization graph near the energy of the first resonace $E_r=1.5388$~\au.
	These results are based on the box size of $L_0=50$~\au ~and the number of basis functions $max(n)=500$, Eq.~\ref{eqbas}. 
	(b) Non-adiabatic coupling elements $\<\psi_1|d/d\eta|\psi_2\>$ corresponding to the avoided crossings. $N_c(\eta)$ are later used in a diabatization procedure
	to construct $2\times 2$ diabatic Hamiltonians corresponding to the avoided crossings.} 
	\mylabel{F2}
\end{figure}

Formally, the complex scaling method corresponds to taking an imaginary value of the scaling parameter $\eta$.
It is thus basic to understand what happens to the spectrum as $\eta$ is taken to the complex plane.
It is known since the stabilization method has been proposed
that each avoided crossing on the real axis is associated with an exceptional (branching) point 
singularity (EP) in the complex plane, see Ref.~\cite{McCurdy1983b}. Two selected avoided crossings and
 EPs are demonstrated for our model in Fig.~3.
EPs often indicate a boundary between different qualitative modes of the system studied~\cite{Bender:1998,Bender:2007,Klaiman:2008}.
The present case is no different.
Before reaching an EP, we find avoided crossings created due to mixing of two adiabatic states, but after the EP there are two
diabatic-like states crossing each other, where one is clearly the resonance while
the other is a detached quasi-bound state. 
Apparently, the EP is associated with a {\it adiabatic to diabatic spectral transition},
which takes place in the complex plane of the scaling parameter $\eta$.
\begin{figure}
	\includegraphics[width=3in]{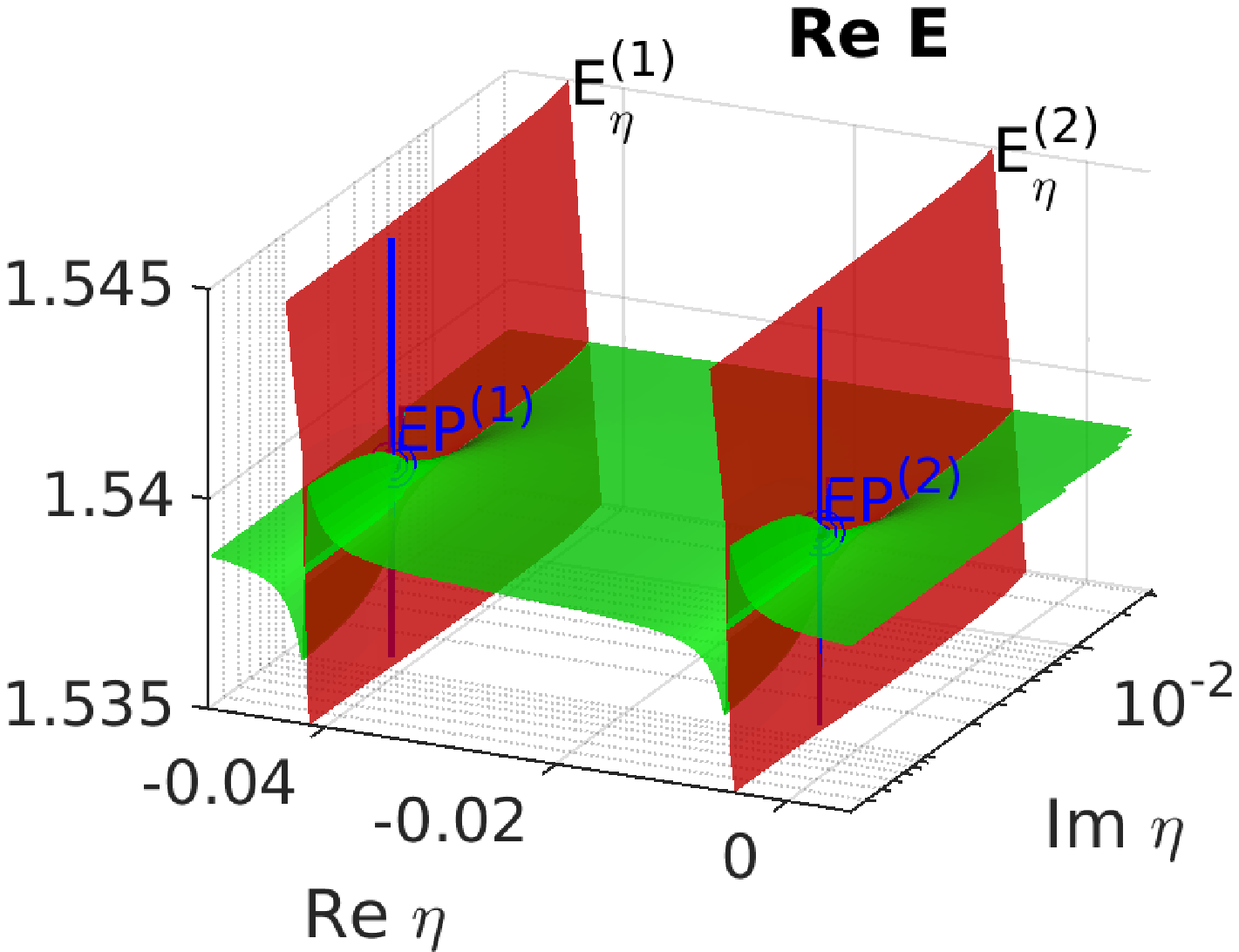}
	\includegraphics[width=3in]{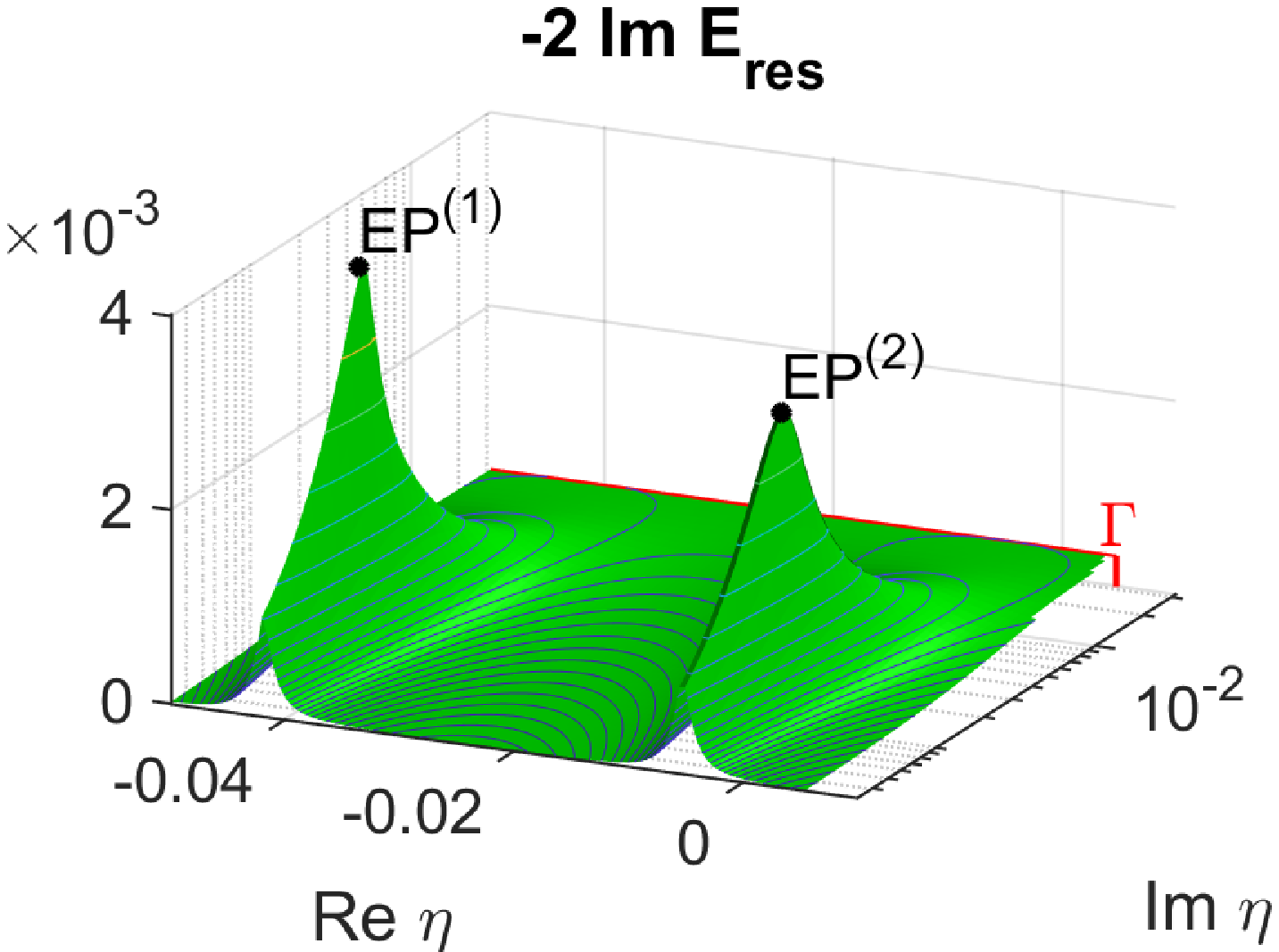}
	\caption{Illustration of the analytical continuation of the real scaling parameter to the complex plane. As an imaginary part is added to the
scaling parameter $\eta$, the energy splitting at the avoided crossings becomes smaller. This trend
continues until the corresponding exceptional points (EP$^{(1)}$, EP$^{(2)}$) are reached.
	Beyond the EPs, the avoided crossings are no more present, instead, the real parts of the energy surfaces for the resonance
	and quasi-continuum cross each other. This phenomenon
	can be interpreted as a natural diabatization at the exceptional point.
	The imaginary part of resonance energy $E_{r}$ is stabilized far in the complex plane of $\eta$ (note the logaritmic scaling of $\im\,\eta$),
	where it corresponds to the physical resonance width $\Gamma$.
	} 
	\mylabel{F3}
\end{figure}

\section{Complex scaling applied ex post to Hermitian spectrum}

\subsection{Basic assumptions}

\noindent
In the second part of the article I will show how all this can be used to calculate the
complex resonance energy. The method starts from the stabilization diagram.
First note that the avoided crossings are relatively far apart,
so that in a good approximation a contribution of only two states can be assumed.

Another important observation is that the EPs are located relatively near the real axis
(the value of $\im\eta$ to find the EP is in fact an order of magnitude smaller then
values of the same quantity where the resonance width has converged to $\Gamma$ (see Fig.~3b)).
Thus the first premise of the new method is that the avoided crossing of the real energies
belongs to the immediate vicinity of the near EP; the EP determines the parameters
of the avoided crossing and vice versa these parameters can be used to determine the EP.

As discussed above, the complex resonance energy is found on the opposite side of
the EP. Not only that, but even at a relatively large distance from it (see Fig.~3b).
Two things are learnt from this circumstance.
First, it is necessary to correctly design the dependence of the quasi-continuum energy
on the scaling parameter $\eta$. For example, if a linear dependence was used as the simplest option,
a linear decrease of the resonance width  instead of its
stabilization with $\im\eta$ would be the wrong result.
Second, the area where the complex resonance energy is stabilized is so distant
that the widths of the continuum states far exceed the separations of the EPs;
thus the area belongs to the common neighborhood and influence of a number of EPs.
%An applicable diabatic Hamiltonian 
%must be composed of many 2$\times$2 Hamiltonians corresponding to
%the avoided crossings on the real axis, which are put together. 

\subsection{From avoided crossings to EPs}
\noindent
Avoided crossings  are found in the stabilization diagram
for certain real values of $\eta=\{\eta_{c1},\eta_{c2}, ...\}$.
Clearly, a diabatic crossing (when no coupling is present)
would occur exactly at the point where the energies
of the resonance $E_r$ and the box state $E_\eta$ are equal.
A diabatic Hamiltonian near the avoided crossing reads such as
\begin{align}
	H=\MM{E_r}{\delta/2}{\delta/2}{E_\eta} .
	\mylabel{eq6}
\end{align}
By using a standard diabatization procedure, it is easy to
fit a particular problem to this formula, where it is found
that $E_\eta$ dependence on $\eta$ is nearly linear:
\begin{align}
	E_\eta = E_r - a\cdot(\eta-\eta_c) ,
	\mylabel{eq2}
\end{align}
whereas the other parameters $E_r$ and $\delta$ are more or less constant within the
range of the crossing. 
The solutions of the diabatic Hamiltonian, Eq.~\ref{eq6}, are given by
\begin{align}
	\epsilon_\pm = E_r + \frac{E_\eta - E_r}{2}
	\[ 1 \pm \sqrt{ 1 + \(\frac{\delta}{E_\eta - E_r}\)^2 } \] .
	\mylabel{eq7}
\end{align}
Clearly, the potential curves are at the closest attachment on the real axis for
$E_\eta=E_r$, where
\begin{align}
	\(\epsilon_+ - \epsilon_-\) \large |_{\eta = \eta_c} = \delta .
	\mylabel{eqdel}
\end{align}
The exceptional point occurs for complex $\eta=\eta_{EP}$, where
\begin{align}
	E_\eta - E_r = - i\delta ,
\end{align}
from where the degenerate complex energy is given by
\begin{align}
	\epsilon_\pm\large|_{\eta=\eta_{EP}} = E_r - \frac{i\delta}{2}.
\end{align}
\begin{figure}
	\includegraphics[width=3in]{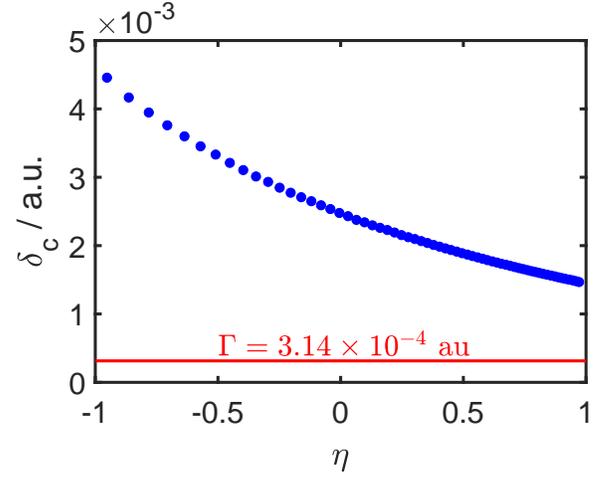}
	\caption{Energy splittings $\delta_c$ for the avoided crossings defined by their positions $\eta=\eta_c$ in the stabilization graph.
		The energy splittings converge to the resonance energy width $\Gamma=3.14\times 10^{-4}$~\au ~in the limit $\eta\to\infty$.
	}
	\mylabel{F2}
\end{figure}

\subsection{Dependence of quasi-continuum on the axis scaling parameter $\eta$\mylabel{Scc}}

\noindent
As long as the box is small, we find out distinct avoided crossings in the stabilization graph,
where the resonance state is represented by a nearly constant (real defined) energy $E_r$.
On the other hand, the intersecting curve of $E_\eta$, which is decreasing with $\eta$,
represents a particle freely moving outside of the potential, within the box. 
We find the dependence 
on $\eta$ using the Schr\"oedinger free particle equation:
\def\T{-\frac{\hbar^2}{2\mu}\frac{\partial^2}{\partial x^2} }
\begin{align}
	\T\psi(x) = E_0 \psi(x),
	\mylabel{sc1}
\end{align}
where upon the scaling the wavefunction is changed such as $\psi(x)\to\psi(xe^\eta)$.
Expectably, this would also bring about the energy change, $E\to E_\eta$:
\begin{align}
	\T \psi(xe^\eta) = E_\eta \psi(xe^\eta).
\end{align}
By changing the variable $x'=xe^\eta$ we get:
\def\T{-\frac{\hbar^2}{2\mu}\frac{\partial^2}{\partial x'^2} }
\begin{align}
	\T \psi(x') = e^{2\eta} E_\eta \psi(x').
	\mylabel{sc3}
\end{align}
By comparing Eqs.~\ref{sc1} and \ref{sc3} we get the dependence of the
free particle states upon the scaling parameter, 
\begin{align}
	E_\eta = E_0\,e^{-2\eta}.
	\mylabel{eq5}
\end{align}

To match the dependence of Eq.~\ref{eq2} with that of
Eq.~\ref{eq5} as close as possible,
I make use of the approximation
\begin{align}
	E_\eta = E_r\[1-\frac{a}{E_r}(\eta-\eta_c)\] \approx E_r\,e^{-\frac{a}{E_r}(\eta-\eta_c)} .
	\mylabel{eq9}
\end{align}
As one would find empirically, the exponent $a/E_r$ is not
given exactly by $2$, see Fig.~\ref{Figexp}, 
calling into question the validity of Eq.~\ref{eq5}.
This discrepancy is perhaps explained by the
influence of non-zero potential even if
the particle is moving outside of the potential trap,
which is the case especially if the box is small (small values of $\eta_c$). 
\begin{figure}
	\includegraphics[width= 3in]{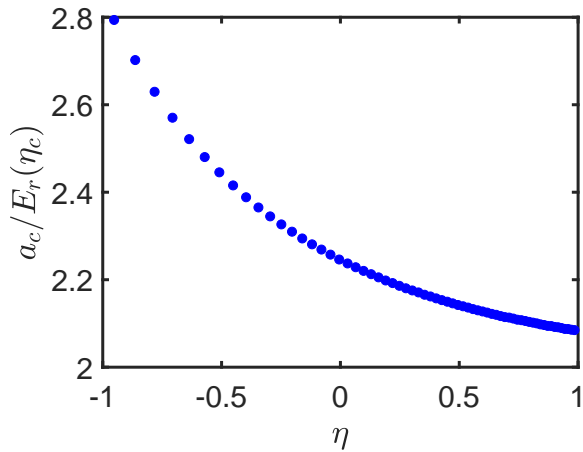}
	\caption{Resonance and quasi-continuum states are diabatized at the avoided crossings, such
	that they cross for $\eta_c$. The diabatic quasi-continuum energy $E_\eta$ is
	given by a linear curve in $\eta$ with the slope of $a_c$ in the short interval of the avoided crossing.
	However, the long range behavior of the quasi-continuum energy
	is exponential such that $E_\eta \propto \exp(-2\eta)$, which applies in the limite of the infinite box size.
	This plot shows that the actual exponent at the avoided crossings is larger then the limiting value of $2$,
	$E_\eta \propto \exp[-a_c/E_r(\eta_c)\cdot (\eta-\eta_c)]$.
	}
	\mylabel{Figexp}
\end{figure}

\subsection{Cooperate remote behavior of exceptional points}

\noindent
As discussed earlier, the EPs are located near the real axis of the scaling parameter
$\eta$ and thus they can be determined by using the avoided crossings in the stabilization
graph.
To find the complex resonance energies, however, it is necessary
to explore the behavior of the complex energies beyond the EPs,
far in the complex plane of the scaling parameter $\eta$.

Let me remind now that some stabilization methods rely on
a very precise and fine fitting of the region near a single EP
to get a good approximation for the distant regions where the
resonance energy gets stabilized~\cite{McCurdy1983b}.
Some other works rely on a fine fitting of the region between the
avoided crossings. As one can see in Fig.~3 that both approaches are
justified due to the principle of
 analytical continuation.

Here we introduce a different approach. While it is based on a physically
justified dependence  for the diabatic states near
avoided crossings, Eq.~\ref{eq9}, the analytical form is too simple to
suffice for an application of the analytical continuation principle.

Clearly, as $\eta$ is taken into complex plane, 
the width of the quasi-continuum states is increased,  Eq.~\ref{eq9}.
One can view these states (taken to the complex plane),
 as a number of energy intervals 
which overlap.
In this picture, many quasi-continuum states overlap
near the resonance energy $E_r$, and therefore are bound to
have some contributions to the resonance.
In order to take into account more states of the quasi-continuum,
it is possible to construct the diabatic Hamiltonian for
several avoided crossings as a single matrix
\begin{align}
	&
	H(\eta) =  \cr
	&\MMF{E_r(\re \eta)}
	{E_{r1}\, e^{-\alpha_1 (\eta-\eta_{c1})}}
	{\dots}
	{E_{r,n}\, e^{-\alpha_n (\eta-\eta_{c,n})}}
	{\delta_{c1}/2}
		{\dots}
		{\delta_{c,n}/2} ,
		\cr
		\mylabel{eqH}
\end{align}
where $\alpha_j$ are defined as
\begin{align}
\alpha_j = \frac{a_{c,j}}{E_{r,j}} .
\end{align}
This Hamiltonian
is based on the 2$\times$2 diabatic Hamiltonians for the individual
avoided crossings on the real axis (Eqs.~\ref{eq6} and \ref{eq9}).
The spectrum of the Hamiltonian $H(\eta)$ corresponds
to that of the usual complex scaled Hamiltonian, see Fig.~\ref{FigCS}.
The difference is that now only one resonance is obtained in the
non-Hermitian spectrum. 
This method can be understood as a {\it complex scaling applied onto the
real spectrum, i.e. ex post} the Hermitian calculation.
\begin{figure}
	\includegraphics[width= 3in]{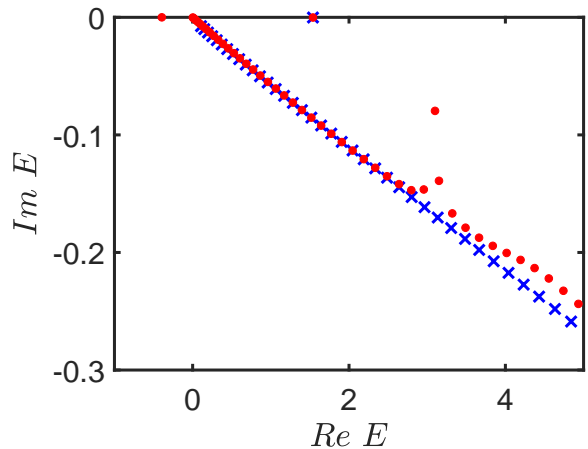}
	\caption{
	This is the spectrum  of the Hamiltonian constructed from several
	avoided crossings on the stabilization graph ('$\times$'), where the complex
	scaling was applied ex post to the quasi-continuum states, see Hamiltonian 
	in Eq.~\ref{eqH}, $\eta=0.025i$.
	It is compared with the calculation where the complex scaling was
	applied directly to the $x$-axis in the Hamiltonian ('$\bullet$').
	The same basis sets were used for both calculations.
	}
	\mylabel{FigCS}
\end{figure}

A sufficient number of the quasi-continuum states must be included in Eq.~\ref{eqH}
to accurately reproduce the resonance energy.
The plots in  Fig.~\ref{FigResEtac} show how the resonance energy is
changed as the quasi-continuum states are added one by one starting
from the avoided crossing for the smallest size of the box $L\approx20$~\au\ 
up to the largest box of $L\approx130$~\au, which correspond to the interval of
the scaling parameter $-1<\eta_c<1$. 

The most significant change of the result
occurs when the states participating on the avoided crossings for the box size of $L\approx 50$~\au\  are included, Fig.~\ref{FigResEtac}.
This value corresponds to the box size when no scaling is used, $\eta=0$. 
These particular states participate on the EPs which are the nearest to the calculated
point in the complex plane, which is defined by $\eta=0.01i$.
\begin{figure}
	\includegraphics[width= 3in]{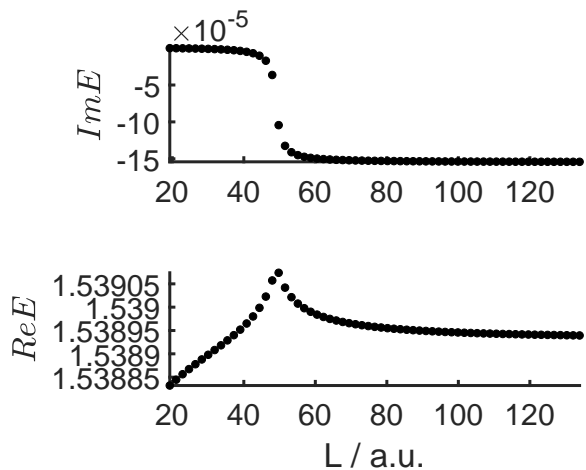}
	\caption{
	Dependence of the resonance energy on the size of the Hamiltonian (Eq.~\ref{eqH}, $\eta=0.01i$),
	as the contributions of subsequent EPs (which correspond to
	the avoided crossings in the stabilization graph) are added.
	The horizontal axis shows the box size at which the
	EP, which was added, is found. So the first point has been obtained
	for a 2$\times$2 Hamiltonian using one EP ($L_c=19.3$~\au), 
	the second for a $3\times3$ Hamiltonian
	using 2 EPs ($L_{c1}=19.3$~\au, $L_{c2}=21.1$~\au), etc..
	}
	\mylabel{FigResEtac}
\end{figure}

A stabilization of the resonance
energy with the increasing complex scaling parameter $\theta\equiv-i\eta$
is demonstrated in Fig.~\ref{FigResTh}, where it is compared with the
result of the usual complex scaling method for the same box size ($L=50$~\au)
and basis set ($N=500$). 

Notably,
the error of the resonance width (imaginary value of its complex energy)
linearly increases with $\theta$ where it should be stabilized according to the
benchmark calculation.
This is a convergence problem, where for larger values of $\theta$, a larger
size of the diabatic Hamiltonian is required, namely it is necessary
to include more avoided crossings corresponding to large values of $\eta_c$.

The real part of the resonance energy is stabilized for large values of $\theta$,
however it includes a constant error. This error
is decreased as more quasi-continuum states are added for the
large box sizes. This indicates that also this error is a matter of convergence.

Should a full convergence be obtained, the avoided crossings for large boxes
are necessary. This in turn requires using more basis functions for the
Hermitian calculations. Using extrapolation to obtain parameters
for the distant avoided crossings ($\eta_c \gg 1$) may help to meet this
requirement in practical applications to quantum chemistry.
\begin{figure}
	(a)\includegraphics[width= 2.9in]{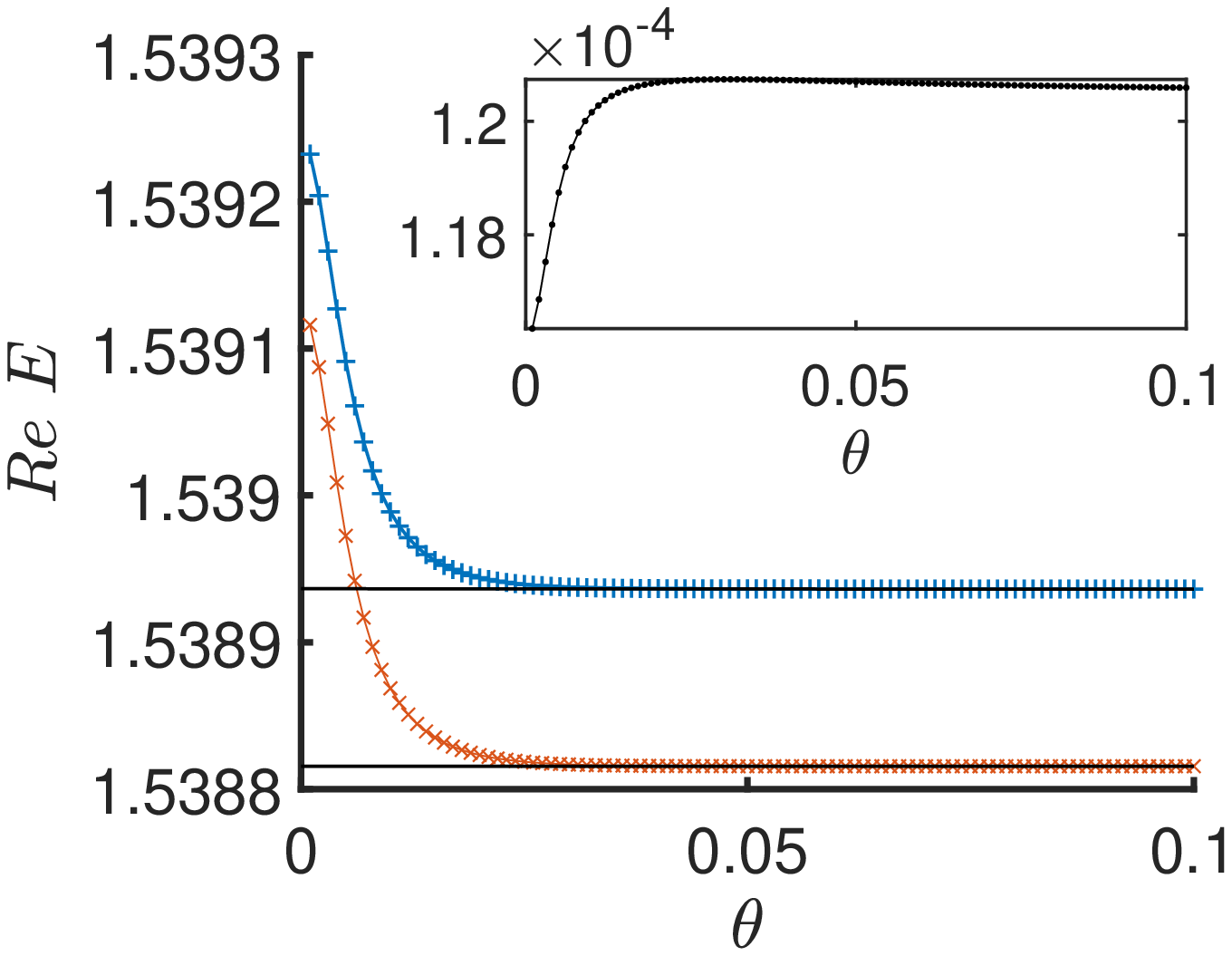}
	(b)\includegraphics[width= 2.9in]{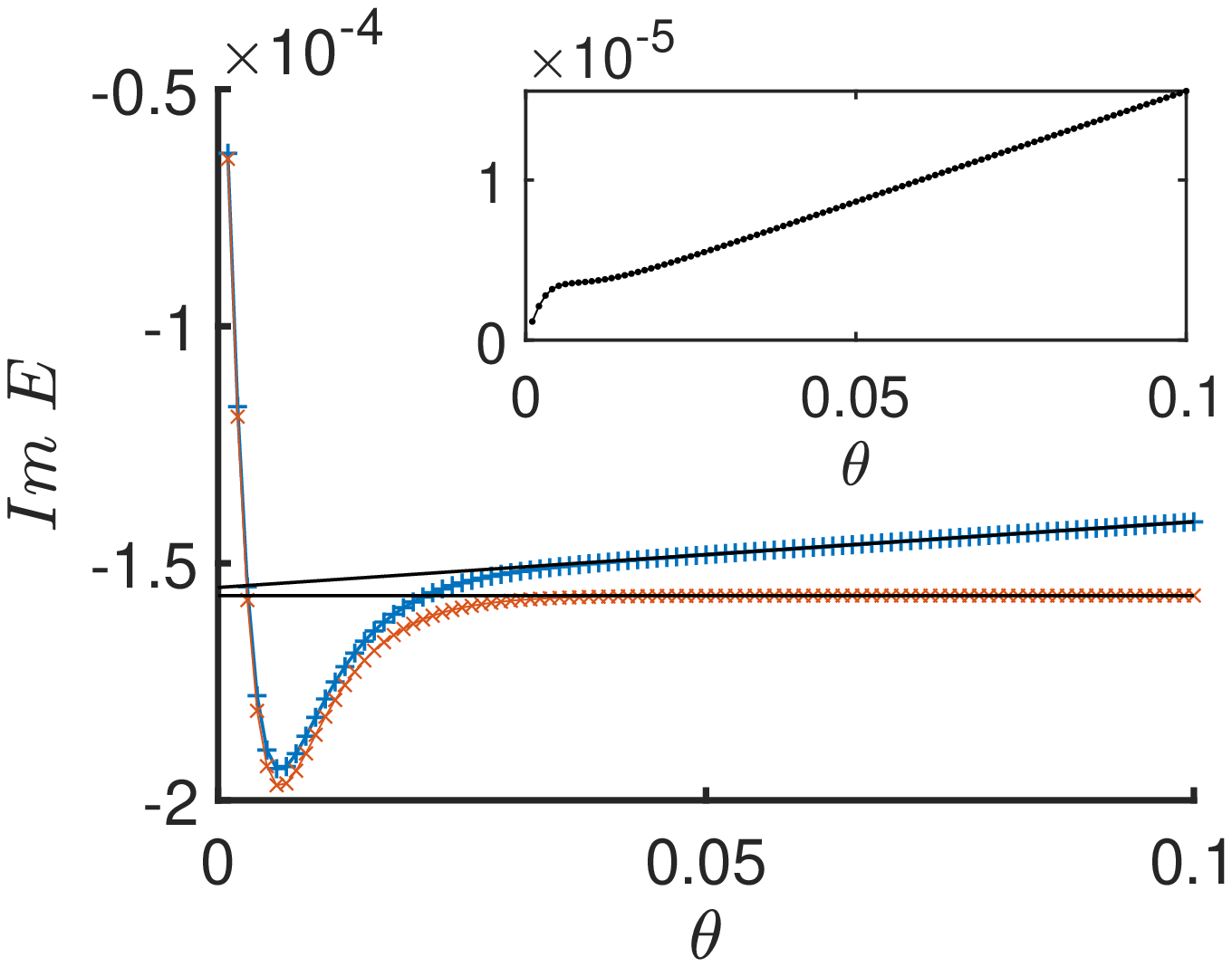}
	\caption{
	Complex resonance energy which has been obtained using the diabatic Hamiltonian (Eq.~\ref{eqH})
	constructed from several EPs corresponding to the avoided crossings
	in the stabilization graph for $-1<\eta_c<1$. The diabatic Hamiltonian
	has been complex scaled by using $\eta=i\theta$, (blue '$+$'). 
	As a benchmark, the result from the usual complex scaling method
	is plotted (red '$\times$').
	The insets show the difference between the approximate and benchmark calculations.
	}
	\mylabel{FigResTh}
\end{figure}

\section{Application for helium doubly excited states}

\subsection{Summary of the new methodology}

\noindent
Let me demonstrate how the described findings can be used as a method for
calculation of atomic resonances.

The proposed method is based on using a large scaling interval.
It is therefore necessarry to choose a basis set which enables
this without a significant precision loss.
Here I show full-CI calculation of doubly excited $2s^2$ state of helium using
exponentially tempered primitive Gaussian basis sets
ExTG5S and ExTG5P optimized for up to four excited bound states
at seven digits of accuracy~\cite{Kapralova-Zdanska:2013a}.
%Such basis sets proved accuracy nearly as high upon direct application
%of complex scaling to obtain complex helium resonances.

In the case of one-dimensional model a standard diabatication procedure, 
based on integrating over the non-adiabatic coupling element, has been used.
Its application to atomic calculations would be cumbersome.
Below I propose a suitable diabatization procedure which avoids calculations
of non-adiabatic coupling elements, being based solely on a precise 
fitting the potential energy curves.

I improve the procedure in several other aspects such as:
(i) I suggest to find a correct analytical
fit for the quasi-continuum by including parts of the potential energy curves
of decoupled quasi-continuum states.
(ii) As in the case of the one-dimensional model potential, also
in atomic application, the complex resonance energy for
large values of $\theta$ sort of drifts out of the correct value, which
requires a backward extrapolation to $\theta\to 0$.
(iii) It is found that the real part of complex resonance
energy is quite underestimated where the error corresponds
to the shift of the ionization potential due to the finite size
of the basis set.

\subsection{Diabatization procedure using ab initio energies\mylabel{SVB}}

\noindent
Let us start our discussion with the $s$-limit, which means
that the atomic basis set is represented only by the $s$-type functions.
The corresponding calculations are very fast and allow us to
calculate a large scale stabilization graph, which is shown
and described in Fig.~\ref{FigStbS}.
\begin{figure}
	\includegraphics[width= 2.9in]{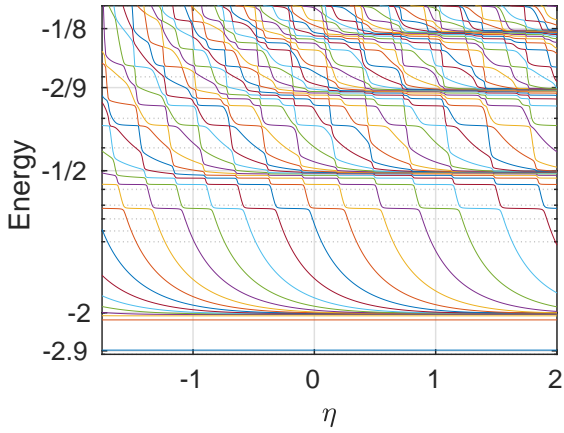}
	\caption{
		A stabilization graph obtained for helium via real scaled Gaussian basis set
		in the $s$-limit.
		All states up to the two-electron ionization limit, $E=0$, are plotted,
		where the energy is shown in the logarithmic scale to comprehend all
		available one-electron ionization thresholds at $-2$~\Eh, $-1/2$~\Eh. etc..
		The ground state level is shown at $-2.9$~\Eh.
		Resonances are manifested by constant energy curves which are disrupted
		with various avoided crossings with the continuum states.
		The discretized continuum states are characterized via exponential decay
		of their energy with the real scaling parameter $\eta$, where they
		converge to the limits associated with the corresponding ionization
		thresholds.
	}
	\mylabel{FigStbS}
\end{figure}

The next goal is to construct the diabatic Hamiltonian defined in Eq.~\ref{eqH}
for a selected helium resonance. We will study the resonance $2s^2$, which
is the lowest metastable state above the first ionization threshold
near the energy $-0.8$~\Eh, see Fig.~\ref{FigStbS}.
The real Hermitian calculations led to thirteen avoided crossings
with different quasi-continuum states as the real scaling parameter
$\eta$ varied in the interval $-1.75\le\eta\le 2$.
 
As the first step we determine two-by-two diabatic Hamiltonians (Eq.~\ref{eq6})
corresponding with the individual avoided crossings.
Let us present here a summary of a robust algorithm
suitable for this task.

At first we select points in the potential energy curves
defined by a confined energy interval near the resonance energy. 
In our case, we used the interval of $\pm 0.1$~\Eh~
around the resonance energy $-0.723$~\Eh, which we could estimate
from the stable parts of the stabilization graph, Fig.~\ref{FigStbS}.
This step allows us split individual avoided crossings which are
processed individually in the next steps.

Real resonance energy $E_r$ and coupling element $\delta$
are obtained in two steps of the predictor-corrector type.

\subsubsection{Predictor}

\noindent
For each avoided crossing we calculate a first estimate of
$\delta$ and $E_{r}$ which define the diabatic Hamiltonian Eq.~\ref{eq6}.
For this sake we use definition of $\delta$ as the minimum energy
split, Eq.~\ref{eqdel}. In practice we use spline interpolation
for the split $(\epsilon_+ - \epsilon_-)$ on the axis of $\eta$
between the known {\it ab initio} points, and find $\delta$
as the interpolated minimum at $\eta = \eta_c$. 

Based on Eq.~\ref{eq7} we get the relation between the sums of
adiabatic and diabatic energies,
\begin{align}
\epsilon_+ + \epsilon_- = E_{r} + E_\eta .
\mylabel{EQ15}
\end{align}
We use the fact that the diabatic energies $E_{r}$ and
$E_\eta$ are equal at the point $\eta=\eta_c$, i.e.,
\begin{align}
E_{r} = \frac{\epsilon_+ + \epsilon_- }
{2} \bigg|_{\eta=\eta_c}\  ,
\end{align}
to get $E_{r}$.
Namely, we calculate the mean adiabatic energy $(\epsilon_+ + \epsilon_-)/2$
for the {\it ab initio} points within the avoided crossing interval,
and then we use the spline interpolant to get the estimate of $E_{r}$ at the
point $\eta_c$ obtained before.

As a matter of fact, the algorithm described here as {\it predictor}
is not robust, as the precision of $\eta_c$ 
largely relies on the spline interpolation. Yet the
above obtained parameters $\delta$ and $E_{r}$ provide
a well needed estimate for the precision
procedure described as {\it corrector}.

\subsubsection{Corrector}

\noindent
We determine the values of $E_{r}$ and $\delta$
in a robust way by fitting them to all {\it ab initio}
points within the avoided crossing.
We use the fact that $E_{r}$ and $\delta$ must represent
constants along the examined interval of $\eta$.
$\delta$ and $E_{r}$ are expressed using the adiabatic energies
such that,
\begin{align}
& \delta = \sqrt{\Delta_\epsilon^2 - 4 (\bar{\epsilon} - E_{r})^2 }, \cr
& E_{r} = sign(E_{r}-\bar \epsilon) \cdot \frac{\sqrt{\Delta_\epsilon^2 - \delta^2}}{2} + \bar \epsilon ,
\end{align}
where we define
\begin{align}
& \bar{\epsilon} = \frac{\epsilon_+ + \epsilon_-}{2},
& \Delta_\epsilon = \epsilon_+ + \epsilon_- \ .
\end{align}
We substitute $\Delta_\epsilon(\eta_k)$
and $\bar \epsilon(\eta_k)$ for each {\it ab initio} point $\eta_k$
within the avoided crossing. 
Using the estimated values of $E_{r}$ and $\delta$
on the right hand sides of the equations
we obtain the values $\delta(\eta_k)$, $E_{r}(\eta_k)$.
These values should be $\eta$ independent supposed that
the values for $\delta$ and $E_{r}$ on the right hand
side were correct.
In reality, they are not, which may be used to get
the right values by imposing the requirement of constancy
in a minimization procedure.
We define a quantity $\sigma$ based on the standard deviation
from the mean values of $\delta(\eta_k)$, $E_{r}(\eta_k)$
such that,
\begin{align}
& \sigma = \sqrt{\sigma_1^2 +\sigma_2^2 }\ , \cr
& \sigma_1^2 = \frac{1}{N}\sum\limits_k \[E_{r}(\eta_k) - \frac{\sum\limits_k{E_{r}(\eta_k)}}{N}\]^2 \ , \cr
& \sigma_2^2 = \frac{1}{N}\sum\limits_k \[\delta(\eta_k) - \frac{\sum\limits_k{\delta(\eta_k)}}{N}\]^2 \ .
\end{align}
Then we find parameters $\delta$ and $E_{r}$ associated
with the minimum value of $\sigma$ using a standard
minimization procedure. The algorithm described
as {\it predictor} is useful for getting an initial estimate for
the numerical calculation.

\subsubsection{Improving ab initio data}

\noindent
Parameter $\delta$ (and
also $\eta_c$, see below)
should form a regular series as obtained for the
set of neighboring avoided crossings. If this is not the
case, calculations must be improved by adding {\it ab initio}
data. In practice, we take the calculated new values of $\eta_c$ (see below)
for which we run additional {\it ab initio} calculations.
After repeating the whole procedure three to four times,
well converged parameters ($E_r$, $\delta$, and $\eta_c$) are finally obtained.

\subsection{Fitting form for quasi-discrete continuum\mylabel{Scont}}

\noindent
Finally, we need to determine the parameters which
characterize the diabatized quasi-continuum state, $\alpha_c$ and $E_0$,
as involved in each avoided crossing.

Let us propose a new form for the continuum states
given by
\begin{align}
&E_{\eta,k} = (E_r-E_{0}) \, e^{-\alpha_{c,k}\,(\eta - \eta_{c,k})-\beta_{c,k}(\eta-\eta_{c,k})^2} \cr
&\quad + E_{0} \ ,
\mylabel{eqc}
\end{align}
which differs from Eq.~\ref{eq9}, which was suggested above for the one-dimensional
testing case, in the following
aspects. 

First, Eq.~\ref{eqc} reflects the fact that the first ionization
continuum is not zero but it is given by the
energy of ground state $He^+$ ion, $E_0=-2$~\Eh.
In fact, due to the variational principle, the energy of the helium ion
for the particular basis set should exceed the infinite basis set
limit of $-2$~\Eh. 
In the case of a Gaussian finite basis set, apparently,
the quasi-continuum state
always ends up below the threshold as it is transformed into 
a highly excited Rydberg state for $\eta\to\infty$, see Fig.~\ref{FigStbS}.
Here, for simplicity, we will use the true
limiting value, $E_0=-2$~\Eh, in Eq.~\ref{eqc}.
Within the fitting procedure used for the quasi-continuum, 
which will be described below,
 there will be a need to exclude the area where 
 the quasi-continuum state has changed
 to the Rydberg state.

The second aspect where Eq.~\ref{eqc} has been modified
from Eq.~\ref{eq9}, is represented by
 the quadratic dependence of the exponent on $\eta$.
In Section~\ref{Scc} we derived the exponential dependence 
for the case of a finite box supposed that the discrete basis set
is infinitely large, where we found also the value of the linear
coefficient, $\alpha_c=2$. 
It was found empirically that when the box size is finite,
$\alpha_c$ deviates from $2$, see Fig.~\ref{Figexp}.
Additionally, the Gaussian basis set used for the present case 
 fills the phase space in a subtle
way which differs from the box basis sets.
The ab initio data displayed in Fig.~\ref{FigStbS}
confirm the general exponential dependence of the energy of the
quasi-continuum, see the curves between the first ionization
threshold (-2~\Eh) and the first resonance
(-0.72~\Eh). Yet, these curves cannot be fitted
precisely enough to a linear exponential, therefore a
quadratic polynomial has been introduced empirically for the exponent.

\subsubsection{Using quasi-discrete continuum outside and within avoided crossings}

\noindent
To obtain the best fit of quasi-continuum, we need a large
portion of the curve defined by ab initio data.
The ab initio curve constitutes from the {\it isolated quasi-continuum}
(see Fig.~\ref{FigStbS} in the interval $-1.75$~\Eh$>E_\eta>-0.8$~\Eh).
Note that we excluded energies below $-1.75$~\Eh.
The reason is that the quasi-continuum states change to Rydberg
states as their energies drop down near the threshold energy.

The next part of the quasi-continuum curve, beyond $E_\eta=0.8$~\Eh,
 is {\it embedded in the avoided crossing}.
 In the area of avoided crossing,
 $E_\eta$ can be determined based on the mean value of the crossing curves,
using the known value for $E_r$, as given in Eq.~\ref{EQ15}.

\subsubsection{Fitting procedure for quasi-discrete continuum}

\noindent
We start by determining the parameter $\eta_c$ in Eq.~\ref{eqc},
for which we use only the part of the quasi-continuum curve 
which was originally embedded in the avoided crossing.
We use a simple parabolic fit for $E_\eta(\eta_k)$ such that
\begin{align}
E_\eta(\eta_c) \approx a_2 \eta^2 + a_1 \eta + a_0,
\end{align}
from which
we determine the value of $\eta_c$ such that
\begin{align}
E_\eta(\eta_c) = E_r .
\end{align}

The next step is to determine $\alpha_c$ and $\beta_c$ in Eq.~\ref{eqc},
for which we use the full quasi-continuum curve composed of the
isolated and embedded parts as discussed above.
We apply the weighted least square fitting of
\begin{align}
f_i\equiv \log \, \frac{E_{\eta}(\eta_i) - E_0}{E_r - E_0}
\end{align}
to the second order polynomial
\begin{align}
f_i \approx -\alpha_c (\eta_i - \eta_c) - \beta_c (\eta_i - \eta_c)^2 \, ,
\end{align}
where the weights are given by,
\begin{align}
w_i = \frac{E_r - E_0}{E_{\eta}(\eta_i) - E_0} \ .
\end{align}

\subsection{Complex scaling -- stabilization for $\theta\gg 0$ and backward extrapolation to $\theta\to0$}

\noindent
At this point we have all parameters needed to construct the
Hamiltonian according to Eq.~\ref{eqH}, having in mind the modification
concerning the $\eta$-dependence of the quasi-continuum, Eq.~\ref{eqc}.
The complex scaled Hamiltonian is obtained by setting
\begin{align}
&\eta = i \theta + \Delta_\eta, 
&0<\theta<\frac{\pi}{4}\ .
\end{align}
$\Delta_\eta$ is a small real parameter which will allow us to
see the dependence of results on a real scaling of the basis set.
By diagonalizing the complex scaled Hamiltonian, we obtain the
typical picture including the rotated quasi-discrete continuum, and
the resonance. Let us discuss the dependence of the obtained complex
resonance energy on the complex scaling parameter $\theta$, which
is displayed in Fig.~\ref{FigCSHe}.

The complex energies demonstrate high instability for $\theta<0.4$,
where they largely depend on $\Delta_\eta$.
This is due to the fact that
$\Delta_\eta$ effectively changes the position with respect
to the branch points along the real axis in Fig.~\ref{F3}.

According to the theory in Section~III, which was also approved
by the one-dimensional numerical experiment, the complex resonance
energy should be constant as $\theta$ is large enough.
In fact, the one-dimensional calculation shows that the imaginary
part of the resonance energy is mildly linearly dependent on $\theta$
is the stable part of the complex plane, however 
the first derivative of this dependence can be pushed
down to zero when increasing the basis set.

Here, Fig.~\ref{FigCSHe}, the stable part is characteristically
almost independent on the real scaling shift $\Delta_\eta$.
However, the complex energy, both real and imaginary components of it,
show a low order polynomial dependence on $\theta$.
We explain this artifact as a result of the finite size of the basis set.
Namely, the only states which depend on the scaling $\eta$ in
the diabatic Hamiltonian, are represented by the quasi-continuum,
but we could see above that the quasi-discrete
continuum has a specific
dependence on $\eta$ which is different from the finite basis set, see Eq.~\ref{eqc}.

In other words, the finite Gaussian basis set, optimized for
calculations of the atomic states as is, deteriorates 
by the application of real scaling which apparently
brings in a systematic error as the scaling parameter
is analytically continued to the complex plane.
This error must be removed by an extrapolation from the
stable region $\theta>0.4$ as if back
to $\theta=0$ as shown by the solid lined
in Fig.~\ref{FigCSHe}.

The complex resonance energy which is obtained for $\Delta_\eta=0$
is given by $E_{2s^2} = (-0.7213  -1.2\times 10^{-3}i)$~\Eh.
The value which is obtained via direct application of the
complex scaling with the same basis set is given by 
$E_{2s^2} = (-0.7228358  -1.199\times 10^{-3}i)$~\Eh, which
also represents the true $s$-limit for this state, see Ref.~\cite{Kapralova-Zdanska:2013}
and references therein.
Clearly, the real scaling method provides a correct
result yet with much larger error bars compared to a direct
application of complex scaling when using the same high 
quality Gaussian basis set.

\begin{figure}
	\includegraphics[width= 2.9in]{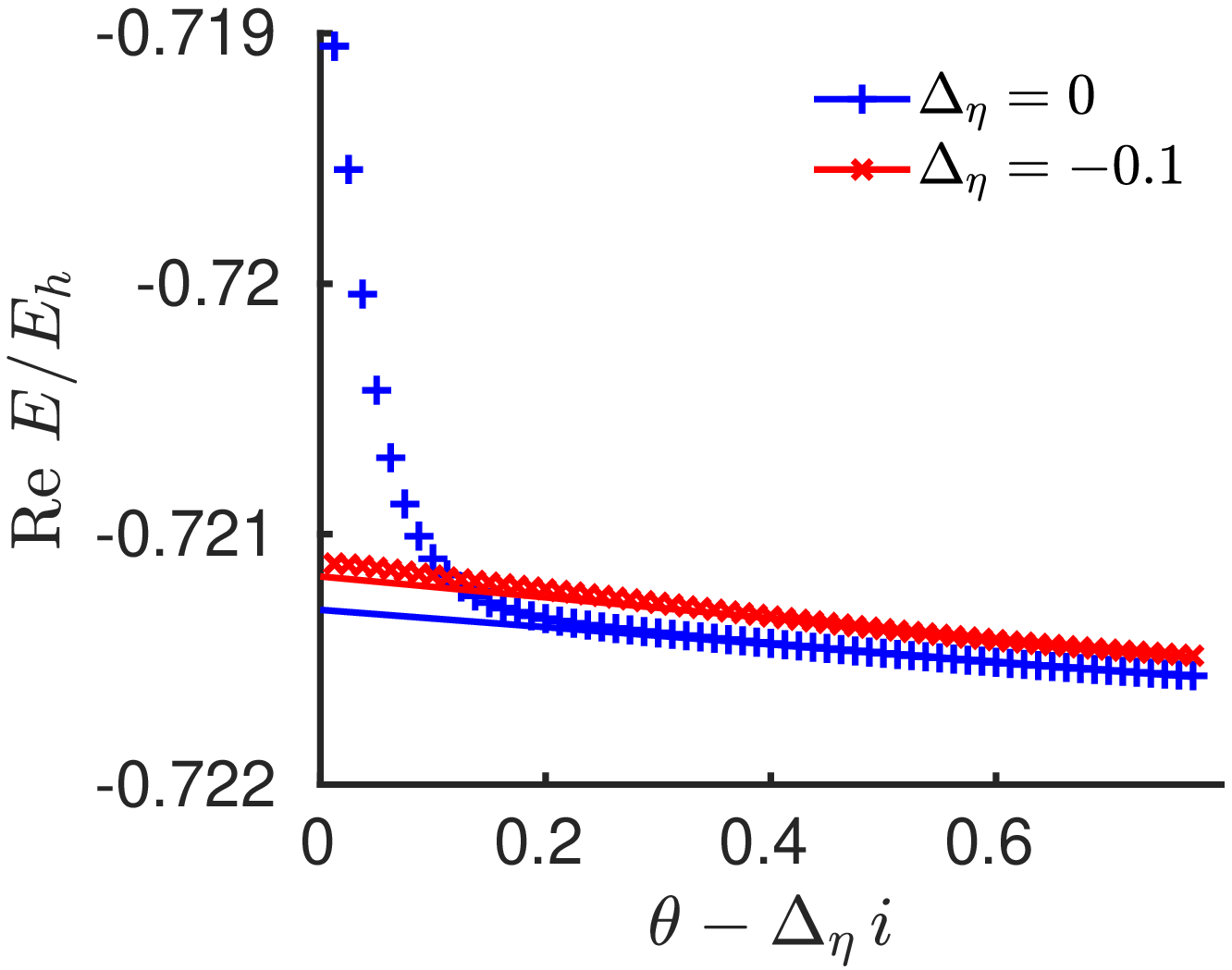}
	\includegraphics[width= 2.9in]{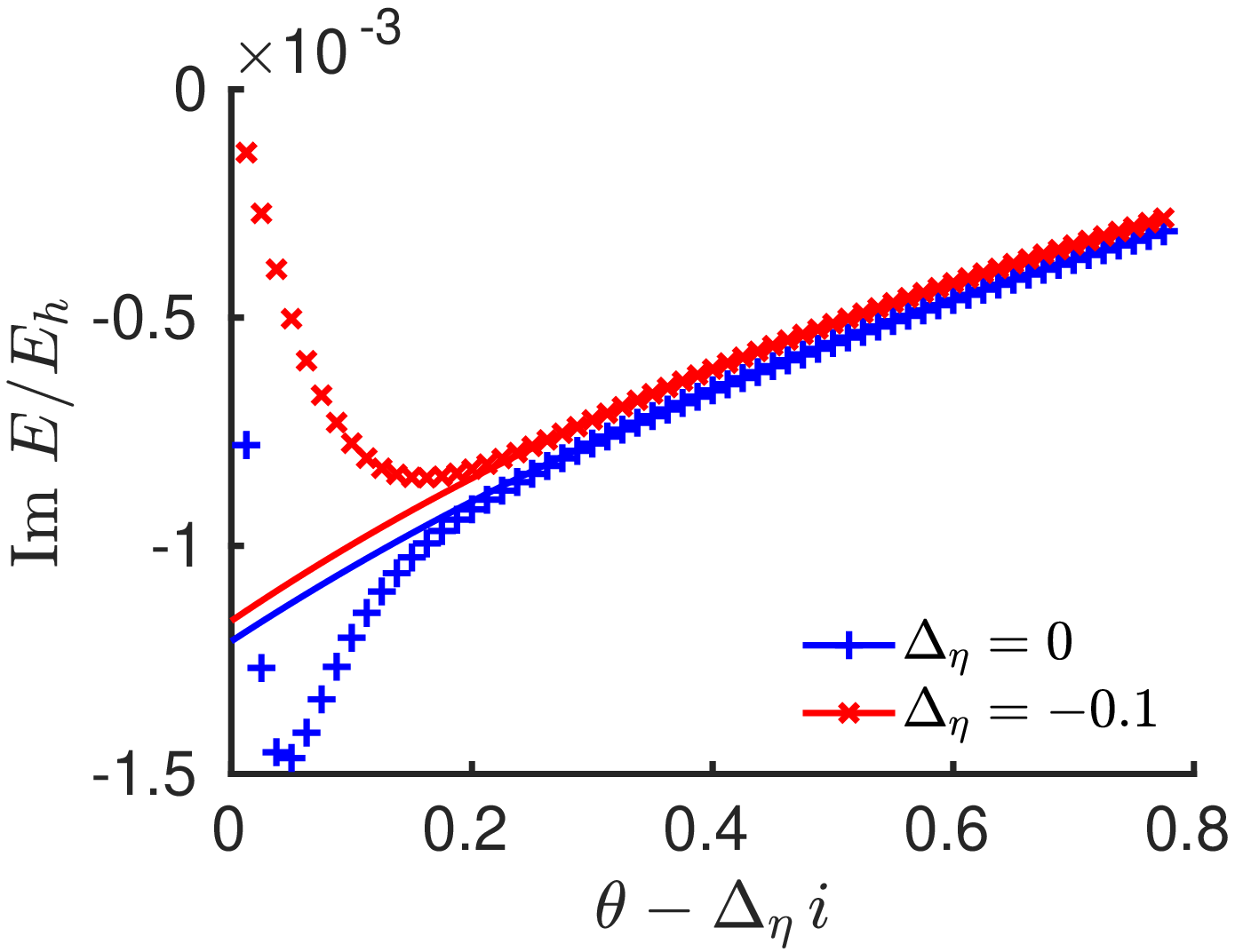}
	\caption{
		Complex resonance energy of the helium $2s^2$ state based
		on the Hermitian calculation of the stabilization graph (Fig.~\ref{FigStbS}).
		The stabilization graph allows us to determine the dependence of
		energies on
		the real parameter $\eta$, which scales the basis set,
		in the form of a diabatic Hamiltonian
		which includes several quasi-continuum states and the resonance (Eq.~\ref{eqH}).
		By bringing the real scaling to the complex plane, $\eta\to i\theta + \Delta_\eta$, we obtain the complex scaled Hamiltonian which is
		diagonalized and the complex resonance energy is obtained (markers $+$,$\times$).
		The real and imaginary parts of the
		resonance energy are greatly unstable for $\theta<0.4$, as they
		pass near branch points, compare Fig.~\ref{F3}.
		The stable part for $\theta>0.4$ is characterized by a
		low order polynomial dependence on $\theta$, which can be used
		to a backward extrapolation to $\theta=0$ as shown by the lines.
		The disturbing fact that the resonance energy is not constant
		even for large values of the complex scaling parameter $\theta$
		is probably explained by the finite size of the basis set, see the text.
		}
	\mylabel{FigCSHe}
\end{figure}

\subsection{Method to increase rotational basis set}

\noindent
Up to now we used only the $s$-type Gaussians to calculate
the $2s^2$ doubly excited resonance of the helium atom.
It is known however that including also the $p$-type functions
leads to a significant increase of the resonance width, see Ref.~\cite{Kapralova-Zdanska:2013} and a notable decrease of the resonance
energy. This is in harmony with the present findings demonstrated
in the comparison of the stabilization graphs for
the $s$ and $p$-limits, Fig.~\ref{FStbP}.
While the decrease of the real part of the resonance energy
with including the $p$ symmetry is obvious,
in particular the increase of the resonance width  can be anticipated from the larger split of the avoided crossings.
\begin{figure}
	\includegraphics[width= 3in]{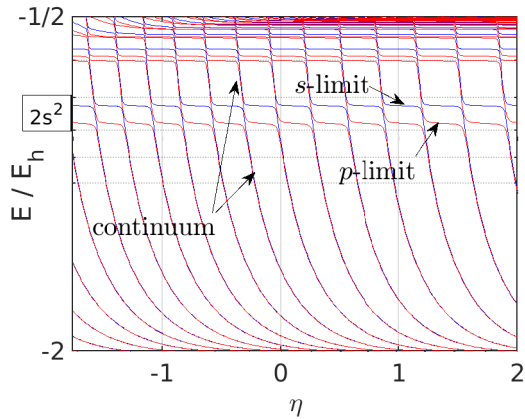}
	\caption{
		Effect of increasing of the rotational basis set 
		-- comparison of
		stabilization graphs in the $s$ and $p$-limits.
		The energy of the $2s^2$ resonance is decreased
		upon adding the $p$ symmetry, while the
		energy split of the avoided crossings is
		increased about twice. 
		On the other hand, the quasi-continuum states
		are almost intact by adding the $p$ symmetry,
		as they correspond to the He$^+$ 1$s$ state
		which has the pure $s$ symmetry.  
	}
\mylabel{FStbP}
\end{figure}

Calculation of massive data when including the $p$ symmetry is still computationally feasible and fast, at least in this system.
However, as the rotational basis set is more increased, the
calculations become costly. 
Therefore we propose using the following computational
 strategy. The positions $\eta_c$ calculated for the $s$ limit
 serve as the initial guess for the higher $p$ limit,
 where we calculate small sets of ab initio points 
near every avoided crossing.
 Then we use again the predictor-corrector
 method described in Section~\ref{SVB} to calculate
 $\delta^{p}$, $E_r^{p}$, and $\eta_c^{p}$, now in the precision
 of the $p$-limit.
 After the parameters in the $p$-limit are known, we
 proceed in the same manner to the $d$-limit
 to get $\delta^{d}$, $E_r^{d}$, and $\eta_c^{d}$, and
 the same could be done also even for higher rotational numbers.

To construct the Hamiltonian Eq.~\ref{eqH}, we need to include the
diagonal terms for the continuum states as well.
We proceed according to Section~\ref{Scont}, where
the continuum is fitted to Eq.~\ref{eqc}
using two parts of the quasi-continuum curve -- 
isolated quasi-continuum, and quasi-continuum
embedded in the avoided crossing.
Now, we use the fact that the isolated part
of the quasi-continuum curve 
is intact by adding the higher rotational symmetries,
see Fig.~\ref{FStbP},
thus the ab initio data obtained for the $s$-limit
can be used for this part.
 
We calculated the complex resonance energy $p$-limit 
using the above indicated algorithm and algorithms
discussed for the $s$-limit. The results are shown
in Fig.~\ref{Figp}.
The calculations were repeated for an interval of the real
basis set scaling $\Delta_\eta$, Fig.~\ref{Figp},
showing a mild dependence on this parameter.
The obtained resonance
 energy is given by $-0.775 -2.37\times 10^{-3}i$~\Eh,
which must be compared with the benchmark value $-0.777296 - 2.332\times 10^{-3}i$~\Eh obtained for the same basis set
when the Hamiltonian is complex scaled directly, Ref.~\cite{Kapralova-Zdanska:2013}.

The error of the present calculation is given by
$+0.002 - 0.00004i$~\Eh, which shows that the error
of the real part is two orders of magnitude
 larger then the error of the imaginary part.
 Note that the same discrepancy occurs for the complex
 resonance energy in the $s$-limit above.
Interestingly, the real shift of the resonance energy 
is comparable with the energy depth (below the
ionization threshold) of the last
Rydberg state which can obtained within the used basis set.
If the energy depth of the last Rydberg state is
substracted from the real part of the resonance energy,
we obtain the ``corrected'' resonance energy 
$-0.77728 -2.37\times 10^{-3}i$ where the error
of the resonance position is given by $0.00002$~\Eh,
now comparable with the precision of the width.
The corrections differ for different
values of $\Delta_\eta$, Fig.~\ref{Figp}, corresponding 
to the last Rydberg state within the real scaled basis set
defined by the scaling parameter $\eta=\Delta_\eta$.

\begin{figure}
	\includegraphics[width= 2.9in]{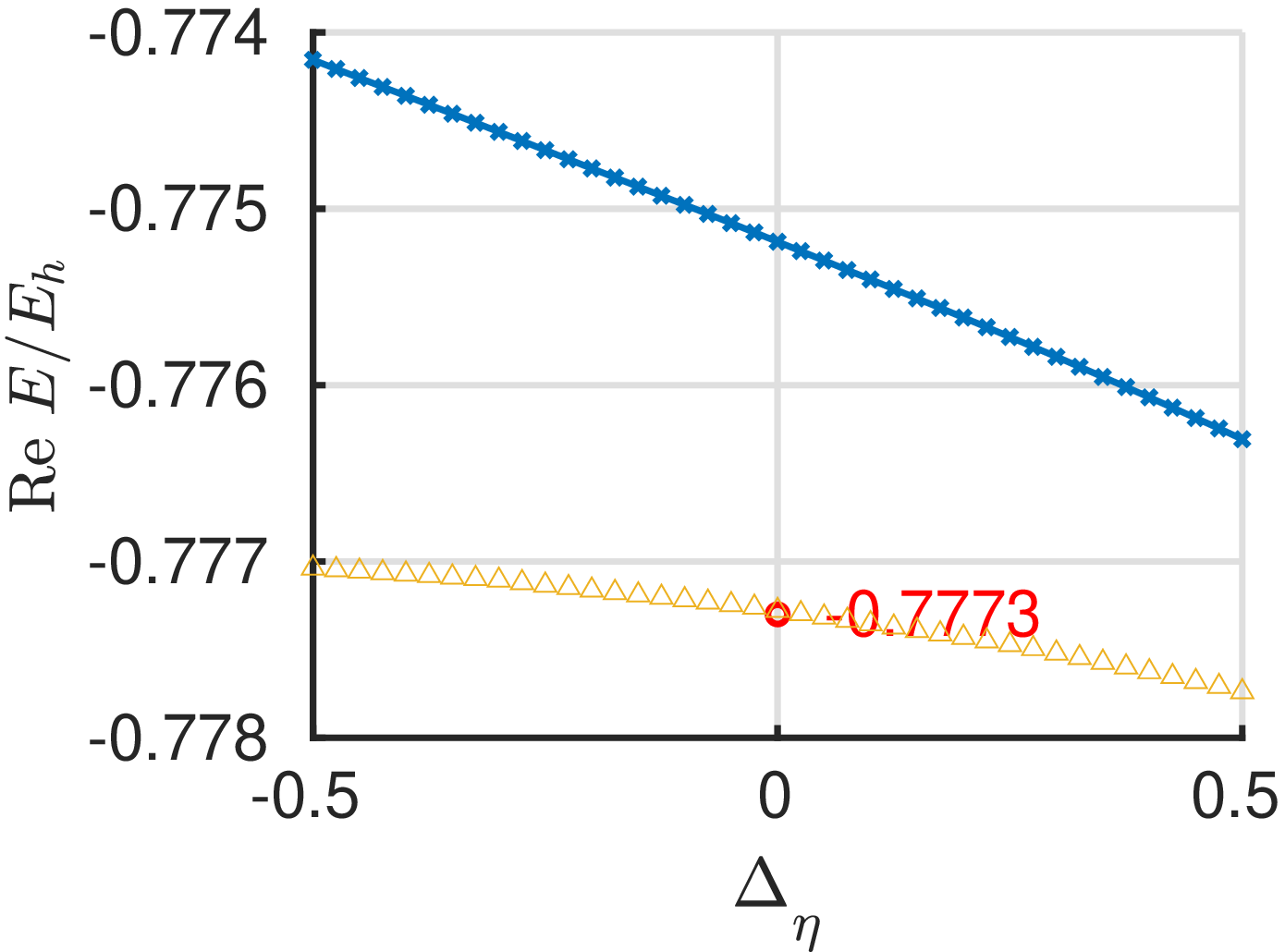}
	\includegraphics[width= 2.9in]{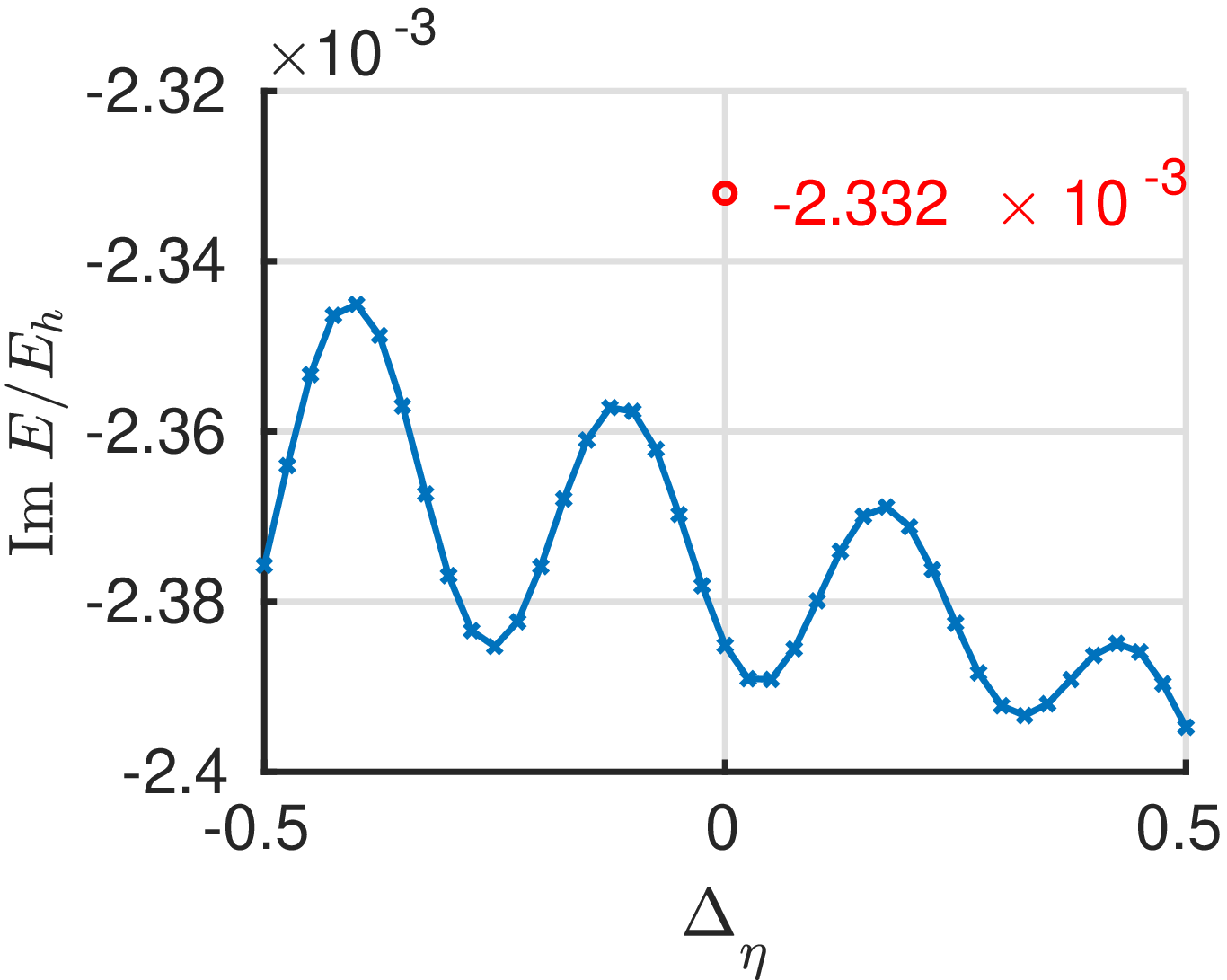}
	\caption{
		Complex energies of the $2s^2$ helium resonance
		obtained for the $p$-limit using real scaled basis set
		calculations are displayed. Cross markers show the results
		which are obtained as indicated in Fig.~\ref{FigCSHe}, i.e.
		via extrapolation to $\theta\to 0$.
		The results depend on the real scaling shift $\Delta_\eta$,
		where the optimal basis set is represented by
		 $\Delta_\eta = 0$ (namely, this point represents
		the original unscaled basis set).
		The circle markers display the results obtained using the
		same basis set when complex scaling is applied to the electronic
		Hamiltonian; they also represent the benchmark values, Ref.~\cite{Kapralova-Zdanska:2013}. 
		The triangles for the real part of the complex resonance
		energy represent a ``corrected'' result, which is obtained
		by adding the energy difference between the true ionization
		potential and the energy of the last bound state 
		within the final basis set.
}
	\mylabel{Figp}
\end{figure}

\section{Conclusions}

\noindent
Hermitian (stabilization) methods for calculations of resonances all boil down to manipulations
with the box size controlled by a ``scaling'' parameter $\eta$. 
The result is represented by the stabilization graph, which includes the
energy spectrum as dependent on the scaling parameter $\eta$.
 The potential energy curves near an energy of a quasi-bound state
 include intervals of $\eta$ where the energy is stable. The intervals
of stability are interupted with avoided crossings due to an interaction with the quasi-free states of the ``box''. 
Each of the avoided crossings corresponds to an exceptional point (EP) in the complex 
plane of the box scaling parameter $\eta$.

I suggest that the EPs can be interpreted as marking a transition between
two qualitatively different descriptions of the problem, where on one side 
the resonance and the quasi-continuum are coupled (which is reflected in the
presense of the avoided crossings), while on the other side of the EP,
the resonance state is decoupled from the quasi-continuum (the potential energy curves of 
the resonance and quasi-continuum cross each other).

I suggest a new method to calculate the complex resonance energy from
the stabilization graph.
Its main idea is represented by appreciating
the fact that the resonance energy is stabilized {\it deep} in the complex plane of $\eta$ where energies of {\it quasi-continuum states} 
 are characterized by  {\it large imaginary parts},
in other words, the quasi-continuum states have large energy widths 
and therefore {\it many} such states
{\it overlap near the position of the resonance}. 
Therefore
{\it all} these states must be included in a diabatic basis set for the resonance.

In accord with this
I proposed a diabatic Hamiltonian which is constructed using
multiple avoided crossings, where each crossing
brings in another quasi-continuum state.
The diabatic Hamiltonian, which is constructed, is parametrized by the 
real scaling parameter $\eta$.
The diabatic Hamiltonian is then analytically continued
to the complex plane through $\eta \to i\theta + \Delta_\eta$.
Diagonalization of such a Hamiltonian leads to the complex energy spectrum 
which includes both the resonance and the rotated quasi-continuum; 
it is in fact directly comparable with the result 
obtained via the usual complex scaling method with the same
basis set.

The new method is developed using a one-dimensional model potential
where a semi-complete basis set is used.
Then the same method is adapted to calculate complex scaled spectrum
of the helium atom, where again a large scale basis set is used.
Let me summarize the pros and cons of the new method.

(i) In contrast to other similar methods based on stabilization
graph, this method does not provide a single resonance energy,
rather it provides a whole {\it part of the complex scaled spectrum}
 including
the resonance {\it plus} several quasi-continuum states.
As such it lends itself directly for calculations of
photoionization resonances (and cross-sections) via methods such as 
$(t,t')$~\cite{ISI:A1994NR28400030,ISI:A1994PV95600060,ISI:A1994NL68500030}
where inclusion of quasi-continuum states is inavoidable.

As another advantage of handling with a portion of the spectrum,
it is well thinkable to extend the diabatic Hamiltonian 
to include even several resonances and bound states.
Such extension would allow to calculate
complex transition dipole moments between resonances and bound states
which are experimentally measurable via absorption Fano profiles~\cite{Pick:2019}.
The transition dipole moments between bound and resonance states
also play a major role in a realization of a recently described
phenomenon of Rabi-to-RAP (Rapid Adiabatic Passage) transition
within dynamical encircling of exceptional point 
in the frequency-laser amplitude plane~\cite{LETTER,PAPER}.

(ii) Interestingly, the application of complex scaling on
the diabatic Hamiltonian which was obtained via fitting on the
real scaled basis set does not inherit numerical problems
which are typical for large values of complex scaling parameter $\theta$
when the system Hamiltonian is complex scaled directly.
Such errors were observed in corresponding complex scaling
calculations of helium atom 
(Ref.~\cite{Kapralova-Zdanska:2013}, Fig.~4)
and were explained in a general detailed study 
of the complex scaling method~\cite{Kapralova-Zdanska:2011}.
Here within the present method, 
a stable complex spectrum is obtained
even up to the limit $\theta\to\pi/4$ (Fig.~\ref{FigCSHe}).

(iii) The present method requires calculation of several
avoided crossings on the stabilization graph which
of course requires a sufficiently large basis set. 
Additionally, a precision
of the obtained results itself
highly depends on using {\it semi-complete
basis sets}. Compared to a direct application of complex scaling
on the Hamiltonian, I obtained two orders of magnitude smaller
precision of the calculated complex resonance energy
for the helium doubly excited resonance $2s^2$
when using the same semi-complete basis set
(yet the precision up to 4$\times 10^{-5}$~\Eh \ has been achieved).

Despite of this disadvantage I beleive that the present
method may be useful especially
for describing laser-atom interactions
typically in situations where a direct application
of complex scaling of Hamiltonian
would represent a technical or numerical problem.

\section*{Acknowledgements}
\noindent
This work was financially supported in parts by 
the Grant Agency of the Czech Republic (Grant No. GA20-21179S) and the Czech Ministry of Education,
Youth and Sports (Grant No. LTT17015).

\end{document}